\documentclass[a4paper,12pt]{article}
\usepackage{amsfonts,slashed}
\usepackage{url}
\usepackage{latexsym}
\usepackage{amsfonts}
\usepackage{epsfig}
\usepackage{latexsym,amssymb}
  \usepackage{amsmath,amssymb,amsthm}
\usepackage{ifpdf}
\ifx\pdfoutput\undefined
   \pdffalse
   \usepackage{cite}
 \else
   \pdfoutput=1
   \pdftrue
  \usepackage[pdftex]{hyperref}
\fi

\numberwithin{equation}{section}
\def\be{\begin{equation}}
\def\ee{\end{equation}}
\def\ba{\begin{array}}
\def\ea{\end{array}}

\newcommand{\bea}{\begin{eqnarray}}
\newcommand{\eea}{\end{eqnarray}}

\def\ii{{\rm i}}

\newcommand{\bbox}{\lower.2ex\hbox{$\Box$}}

\def\bfone{\relax{\rm 1\kern-.35em 1}}
\def\bfzero{\relax{\rm I\kern-.18em 0}}
\newcommand{\so}{\mathfrak{ {}so}}
\newcommand{\osp}{\mathfrak{ {}osp}}

\begin{document}
\numberwithin{equation}{section}

\begin{center}
{\bf\LARGE More on the Hidden Symmetries of 11D Supergravity} \\
\vskip 2 cm
{\bf \large Laura Andrianopoli$^{1,2}$,  Riccardo D'Auria$^{1}$
 and Lucrezia Ravera$^{1,2}$}
\vskip 8mm
 \end{center}
\noindent {\small $^1$ DISAT, Politecnico di Torino, Corso Duca
    degli Abruzzi 24, I-10129 Turin, Italy\\
    $^{2}$  \it Istituto Nazionale di
    Fisica Nucleare (INFN) Sezione di Torino, Italy
}

\vskip 2 cm
\begin{center}
{\small {\bf Abstract}}
\end{center}
\vskip 1 cm
In this paper we clarify the relations occurring among  the $\osp(1|32)$ algebra, the $M$-algebra and the  hidden superalgebra underlying the Free Differential Algebra of D=11 supergravity (to which we will refer as DF-algebra) that was introduced in the literature by D'Auria and Fr\`e in 1981 and is actually a (Lorentz valued) central extension of the $M$-algebra including a nilpotent spinor generator, $Q'$.
We focus in particular on the 4-form cohomology in 11D superspace of the supergravity theory, strictly related to the presence in the theory of a 3-form $A^{(3)}$. Once formulated in terms of its hidden superalgebra of 1-forms, we find that $A^{(3)}$  can be decomposed into the sum of two parts having different group-theoretical meaning: One of them allows to reproduce the FDA of the 11D Supergravity due to non-trivial contributions   to the 4-form cohomology in superspace, while the second one does not contribute to the 4-form cohomology, being a closed 3-form in the vacuum,  defining however a  one parameter family of trilinear forms  invariant under a symmetry algebra related to $\mathfrak{osp}(1|32)$ by redefining the spin connection and adding a new Maurer-Cartan equation.

We further discuss about the crucial role played by the $1$-form  spinor $\eta$ (dual to the nilpotent generator $Q'$) for the $4$-form cohomology of the eleven dimensional theory on superspace.

\vfill
\noindent {\small{\it
    E-mail:  \\
{\tt laura.andrianopoli@polito.it}; \\
{\tt riccardo.dauria@polito.it}; \\
{\tt lucrezia.ravera@polito.it}}}
   \eject

\section{Introduction}

The $\osp(1|32)$ algebra is the most general simple superalgebra involving a fermionic generator with 32 components, $Q_\alpha$,
$\alpha =1,\cdots ,32$. This is also the dimension of the fermionic generator of eleven dimensional supergravity (11D supergravity in the following). It is then natural that, already from the first construction of 11D supergravity in \cite{Cremmer:1978km}, it was conjectured that $\osp(1|32)$ should play a role in the algebraic structure of supergravity, somehow underlying, at least in some contracted version, the eleven dimensional theory. They are however quite different:
 The 11D supergravity \cite{Cremmer:1978km} contains, besides the super Poincar\'{e} fields given by the Lorentz spin connection $\omega^{ab}$ and the supervielbein $(V^a,\psi^\alpha)$, $a=0,1,\cdots ,10$, also a 3-form $A^{(3)}$, satisfying, in the superspace vacuum:
\begin{equation}\label{solito}
d A^{(3)}-\frac12 \bar\psi\wedge\Gamma_{ab}\psi\wedge V^a\wedge V^b=0,
\end{equation}
whose closure relies on 3-fermion 1-forms Fierz identities in superspace. As such, this theory is not based on a superalgebra, but instead on a Free Differential Algebra (FDA in the following) on the superspace spanned by the supervielbein. On the other hand, the fields of the $\osp(1|32)$ algebra are 1-forms dual to generators which include, besides the AdS generators $J_{ab}$ and $P_a$ and the supersymmetry charge $Q_\alpha$, also an extra generator $Z_{a_1\cdots a_5}$ carrying five antisymmetrized Lorentz indices $a=0,1,\cdots,10$, its dual being a five-indexed antisymmetric Lorentz 1-form $B^{a_1\cdots a_5}$.

In the sequel we shall mainly use the dual form of Lie (super-) algebras, written in terms of the Maurer-Cartan equations among 1-forms dual to the commutators of the Lie Algebra generators.
 In the case of $\osp(1|32)$, the relevant set of 1-forms is $\sigma^\Lambda\equiv\{\psi^\alpha,\omega^{ab}, V^a, B^{a_1\cdots a_5}\}$,
dual to the $\osp(1|32)$ generators $T_\Lambda\equiv \{Q_\alpha,J_{ab}, P_a, Z_{a_1\cdot a_5}\}$, respectively. \footnote{They satisfy $\sigma^\Lambda(T_\Sigma)=\delta^\Lambda_\Sigma$, being $\delta^\Lambda_\Sigma$ properly antisymmetrized when $\Lambda$ and $\Sigma$ are sets of antisymmetric Lorentz indices; $d\sigma^\Lambda(T_\Sigma,T_\Gamma)= -\frac 12 C^\Lambda_{\ \Sigma\Gamma}$, $C^\Lambda_{\ \Sigma\Gamma}$ being the structure constants of the Lie super algebra.}
The explicit form of the Maurer-Cartan equations for $\osp(1|32)$ Lie superalgebra, once decomposed in terms of its subalgebra $\so(1,10)$, is:
\begin{equation}
\label{osp32}
\begin{aligned}
d\omega^{ab}& -\omega^{ac}\wedge \omega_{c}^{\ b} + e^2 V^a \wedge V^b+ \frac{e^2}{4 !}B^{a b_1 \ldots b_4}\wedge B^b_{\;\; b_1 \ldots b_4} +\frac{e}{2}\bar{\psi}\wedge \Gamma^{ab}\psi=0 , \\
DV^{a}&-  \frac{e}{2\cdot (5!)^2}\epsilon^{a b_1 \ldots b_5 c_1 \ldots c_5}B_{b_1 \ldots b_5}\wedge B_{c_1 \ldots c_5}-   \frac{\rm i}{2}\bar{\psi} \wedge \Gamma^a \psi =0,
\\D B^{a_1 \ldots a_5} &- \frac{e}{5!}\epsilon^{a_1 \ldots a_5 b_1 \ldots b_6}B_{b_1 \ldots b_5}\wedge V_{b_6}+ \\
& +\frac{5 e}{6!}\epsilon^{a_1 \ldots a_5 b_1 \ldots b_6}B^{c_1 c_2}_{\;\;\;\;\;\;\; b_1 b_2 b_3}\wedge B_{c_1 c_2 b_4 b_5 b_6}- \frac{\rm i}{2}\bar{\psi}\wedge \Gamma^{a_1 \ldots a_5}\psi =0 ,\\
D \psi &- \frac{\rm i}{2} e \Gamma_a \psi \wedge V^a - \frac{\rm i}{2 \cdot 5!} e \Gamma_{a_1 \ldots a_5}\psi \wedge B^{a_1 \ldots a_5}=0,
\end{aligned}
\end{equation}
where $D= D(\omega)$ is the Lorentz covariant derivative and $e$ is a (dimensionful) constant. \footnote{In (\ref{osp32}) we are considering dimensionful 1-form generators. Precisely, the bosonic 1-forms $V^a, B^{a_1\cdots a_5}$ carry length dimension one, the gravitino 1-form $\psi$ has length dimension $1/2$, while the Lorentz spin connection $\omega^{ab}$ is adimensional. As a consequence, the parameter $e$ has dimension $-1$ and can be thought as proportional to (the square root of) a cosmological constant.}

Let us note that the very presence of the generator $B^{a_1\cdots a_5}$ in the simple superalgebra (\ref{osp32}) does not allow to interpret a theory based on such an algebra as a theory on ordinary superspace, whose cotangent space is spanned by the supervielbein $(V^a, \psi$), with Lorentz spin connection $\omega^{ab}$. To allow such an interpretation, the Lorentz covariant derivative of the 1-form fields should be expressed only in terms of 2-forms bilinears of the supervielbein.
This is not possible for the Lie supergroup manifold $OSp(1|32)$, unless one would enlarge the notion of superspace by including the 1-form $B^{a_1\cdots a_5}$ as an extra bosonic cotangent vector, playing the role of a ``dual vielbein''.

\vskip 5mm
The aim of the present paper is to investigate the role played by $\osp(1|32)$ on the FDA of 11D supergravity, and to clarify the analogies and differences between the two algebraic structures.
Referring to what we have discussed so far, the comparison between 11D supergravity and a theory based on the Lie superalgebra $\osp(1|32)$ could be summarized as follows: On  one hand, we have a theory well defined on superspace but involving a 3-form, and therefore based on an algebraic structure which is not associated with a Lie superalgebra, but with a FDA; On the other hand, we have an algebraic structure corresponding to a Lie superalgebra, $\osp(1|32)$, which is however hardly associable to a theory on ordinary superspace, since it defines the tangent space to a Lie supergroup manifold corresponding to an \emph{enlarged superspace} generated by $\{\psi^\alpha,V^a,B^{a_1\cdots a_5}\}$.

A Lie superalgebra of 1-forms leaving invariant 11D supergravity and reproducing, through a procedure that will be shortly reviewed in the following, the FDA \emph{on ordinary 11D superspace}, was introduced already in $1981$ in \cite{D'Auria:1982nx}.
A subalgebra of the superalgebra given in \cite{D'Auria:1982nx} was discovered some years later. It includes a 1-form generator carrying five antisymmetrized Lorentz indices, $B^{a_1\cdots a_5}$, together with another 1-form generator, $B^{ab}$. They are the 1-forms dual to the central generators of a central extension of the supersymmetry algebra including, besides the Poincar\'e algebra, the anticommutator
\begin{eqnarray}
\left\{ Q ,\bar Q \right\} &=&- \left[\ii  \left(C
\Gamma ^{a} \right)  P_{a}+\frac{1}{2}\left( C \Gamma
^{ab}\right) Z_{ab}
+\frac{\ii}{5!}\left( C \Gamma ^{a_1\cdots a_5} \right)Z_{a_1\cdots a_5}\right]\, ,\label{bla}
\end{eqnarray}
where $Z_{ab}$, and $Z_{a_1\cdots a_5}$ are Lorentz-valued central charges.\footnote{Here and in the following the term ``central'' for the charges $Z_{ab}$, $Z_{a_1\cdots a_5}$ and for the spinorial charge $Q'$ that will be introduced later, refers to their commutators with all the generators apart from the Lorentz generator $J_{ab}$; The commutation relations with it are obviously dictated by their Lorentz index structure.}
The super Lie algebra (\ref{bla}) was named $M$-algebra \cite{deAzcarraga:1989mza, Sezgin:1996cj, Townsend:1997wg, Hassaine:2003vq, Hassaine:2004pp}. It is commonly considered as the super Lie algebra underlying $M$-theory \cite{Schwarz:1995jq,Duff:1996aw,Townsend:1996xj} in its low energy limit, corresponding to 11D supergravity in the presence of non-trivial $M$-brane sources \cite{Bergshoeff:1987cm, Duff:1987bx, Bergshoeff:1987qx, Achucarro:1987nc, Townsend:1995kk,Townsend:1995gp}.
The algebra (\ref{bla}) generalizes to 11D supergravity (and in fact, by dimensional reduction, to all supergravities in dimensions higher than four) the topological notion of central extension of the supersymmetry algebra introduced in \cite{Witten:1978mh}, as it encodes the on-shell duality symmetries of string and $M$-theory \cite{Duff:1990hn,Hull:1994ys, Witten:1995ex,Duff:1995wd,Becker:1995kb}.

A field theory based on the {$M$-algebra} (\ref{bla}), however, is naturally described on an \emph{enlarged superspace} spanning, besides the gravitino 1-form, also the bosonic fields $\{V^a, B^{ab}, B^{a_1\cdots a_5} \}$.
If we hold on the idea that the low energy limit of $M$-theory should be based on the same ordinary superspace, spanned by the supervielbein $(V^a,\psi)$, as in 11D supergravity, then the $M$-algebra cannot be the final answer, since its generators
are not sufficient to reproduce the FDA on which 11D supergravity is based.

This issue was raised already in \cite{D'Auria:1982nx}, and solved by further enlarging the set of generators with the inclusion of an extra nilpotent spinor generator.
Indeed, as shown in \cite{D'Auria:1982nx}, the equivalence of a super-Lie algebra with the 11D FDA on superspace (and therefore to the Cremmer, Julia, Scherk theory \cite{Cremmer:1978km}) can be proven if one is able to express the 3-form $A^{(3)}$ as a trilinear polynomial of 1-forms $\sigma^\Lambda$, namely $A^{(3)}= A^{(3)}(\sigma^\Lambda)$, in such a way that $A^{(3)}(\sigma^\Lambda)$ still satisfies the FDA on superspace, eq. (\ref{solito}):
\begin{equation}\label{solitosigma}
d A^{(3)}(\sigma^\Lambda)=\frac12 \bar\psi\wedge\Gamma_{ab}\psi\wedge V^a\wedge V^b\,.
\end{equation}
As shown in \cite{D'Auria:1982nx}, this is only possible if the $\sigma^\Lambda$ (and their dual $T_\Lambda$) close a Lie superalgebra,
 called D'Auria-Fr\'e algebra (DF-algebra in the following), containing the $M$-algebra (\ref{bla}) as a subalgebra, but \emph{ including also a nilpotent fermionic generator $Q'$}, satisfying $Q'^2=0$, dual to a spinor 1-form $\eta$, whose contribution to the DF-algebra Maurer-Cartan equations is:
\begin{equation}\label{deta}
  D\eta=\ii E_1 \Gamma_a \psi \wedge V^a + E_2 \Gamma_{ab}\psi \wedge B^{ab}+ \ii E_3 \Gamma_{a_1\cdots a_5}\psi \wedge B^{a_1\cdots a_5}\,.
\end{equation}
Here, $E_1,E_2,E_3$ are real constant parameters satisfying a suitable linear relation necessary for the closure $D^2\eta =0$ (see equation (\ref{relat}) in the following). Actually, in \cite{D'Auria:1982nx} only two particular sets of $\{E_i\}$ were considered, but later in \cite{Bandos:2004xw} it was shown that in fact a one-parameter family of $\{E_i\}$ can be considered, corresponding to all possible solutions to (\ref{relat}) after choosing the normalization of $\eta$. We will come back to this point in Section \ref{relating}.

In other words, the DF-superalgebra underlies the formulation of the 11D FDA on superspace (and therefore the 11D theory on space-time introduced in \cite{Cremmer:1978km}) once the 3-form is expressed in terms of 1-form generators \emph{including also the $\eta$ 1-form}.

 As it was shown in reference \cite{Andrianopoli:2016osu},  this  in turn implies that the group manifold generated by the DF-superalgebra has a fiber-bundle structure whose base space is ordinary superspace,
while the fiber is spanned, besides the Lorentz spin connection $\omega^{ab}$, also by the bosonic 1-form generators $B^{ab}, B^{a_1,\dots a_5}$.
 In particular, the extra nilpotent generator $Q'$, dual to the 1-form $\eta$, allows to consider the extra 1-forms $B^{ab}$ and $B^{a_1\cdots a_5}$ as gauge fields in ordinary superspace instead of  additional vielbeins of an enlarged superspace. This is due to the dynamical cancellation of their unphysical contributions to the supersymmetry and gauge transformations with the supersymmetry and gauge transformation of $\eta$, according to (\ref{deta}). As observed in \cite{Andrianopoli:2016osu}, all the above procedure of enlarging the field space to recover a well defined description of the physical degrees of freedom is strongly reminiscent of the BRST-procedure, and the behavior of $\eta$ is such that it can be actually thought of as a ghost for the 3-form gauge symmetry, when the 3-form is parametrized in terms of 1-forms.

 The DF-algebra of \cite{D'Auria:1982nx} raised recently a certain interest in the mathematical-physicists community, due to the fact that it can be reformulated in terms of $\mathcal{L}_n \subset \mathcal{L}_\infty$ algebras, or strong homotopy Lie algebras. A comprehensive reference to this approach can be found in \cite{Sati:2015yda, stronghom}.

During the years, many attempts have been made to relate the $\osp(1|32)$ superalgebra to the full DF- or to its $M$-subalgebra, and to 11D supergravity, see in particular \cite{Castellani:1982kd}.
Furthermore, in Refs. \cite{deAzcarraga:2002xi, deAzcarraga:2004zj, deAzcarraga:2007et}, the authors discussed the precise relation of the M-algebra to $\osp(1|32)$. In particular, in \cite{deAzcarraga:2002xi} the general theory of expansions of Lie algebras was introduced, and  applied to the above problem: It was shown that the $M$-algebra can be found as  an expansion of $\osp(1|32)(2, 1, 2)$ (this was further explained in \cite{deAzcarraga:2004zj}). Then, in \cite{deAzcarraga:2007et}   the possibility of an ``enlarged superspace variables/fields correspondence principle in $M$-theory" was discussed.
Important contributions to the relations among the $\osp(1|32)$ superalgebra, the full DF- or to its $M$-subalgebra, and 11D supergravity were also given in \cite{Hassaine:2003vq, Hassaine:2004pp, Bandos:2004xw, Troncoso:1997va, Horava:1997dd, Troncoso:1998ng, Zanelli:2005sa, Izaurieta:2006zz, Izaurieta:2011fr}, mainly in the construction of a Chern-Simons 11D supergravity based on the supergroup $OSp(1|32)$.

In the rest of the paper we are going to show that in fact the DF-algebra, which accounts for the non-trivial 4-form cohomology of 11D supergravity, cannot be found as a contraction from $\osp(1|32)$.
More precisely, we are going to show that the 3-form of 11D supergravity, once parametrized in terms of 1-forms of the DF-algebra, may be decomposed in the sum of two terms:
$$A^{(3)}(\sigma^\Lambda) = A^{(3)}_{(0)} + \alpha A^{(3)}_{(e)}\, ,$$
$\alpha$ being a free parameter. The contribution $A^{(3)}_{(0)} $ explicitly breaks $\osp(1|32)$, giving however the non-trivial contribution to the 4-form cohomology in superspace, while $A^{(3)}_{(e)}$ is a 3-cocycle  of the FDA enjoying invariance under a supergroup related, as it will be explained in Section \ref{unusual}, to $OSp(1|32)$ by redefining the spin
connection and adding a new Maurer-Cartan equation. It is actually the only contribution in $A^{(3)}(\sigma^\Lambda)$
  depending on the 1-form $B^{a_1\cdots a_5}$.

\section{Torsion deforming $\osp(1|32)$}\label{unusual}

The aim of this section is to give the precise relation of
 the $\osp(1|32)$ algebra with the M-algebra, in such a way to be able to compare it, in the following section,
with the DF-algebra and with its $M$-(sub)algebra. 
This should in particular allow to overcome a possible obstruction, due the presence in the $M$-algebra of the 1-form generator $B^{ab}$ associated to the central charge $Z_{ab}$, while no such generator appears in the $\osp(1|32)$ Maurer-Cartan equations (\ref{osp32}).

This problem can be easily overcome by exploiting the freedom of redefining the
 Lorentz spin connection in the $\osp(1|32)$ algebra (\ref{osp32}) by the addition of an antisymmetric tensor 1-form $B^{ab}$ (carrying length dimension 1) as follows:
\begin{equation}\label{newomega}
  \omega^{ab}\rightarrow \omega^{ab} + eB^{ab}\equiv\hat\omega^{ab}\,,
\end{equation}
where $e$ is a dimensionful parameter with length dimension $-1$ (it can then be identified with the one already present in the $\osp(1|32)$ algebra as written in (\ref{osp32})). The discussion presented here essentially follows results obtained in \cite{Castellani:1982kd}.

Note that such redefinition is always possible, and it implies a change of the torsion 2-form.
After this redefinition of the spin connection, renaming $\hat \omega \Rightarrow \omega$ eqs. (\ref{osp32}) take the following form
\begin{equation}\label{osp32'}
\begin{aligned}
& d\omega^{ab} -\omega^{ac}\wedge \omega_{c}^{\ b} - eDB^{ab} - e^2 B^{ac}\wedge B_{c}^{\ b}  + e^2 V^a \wedge V^b+ \\
& + \frac{e^2}{4 !}B^{a b_1 \ldots b_4}\wedge B^b_{\;\; b_1 \ldots b_4} +\frac{e}{2}\bar{\psi}\wedge \Gamma^{ab}\psi=0 \, , \\
& DV^{a} +e B^{ab}\wedge V_b -  \frac{e}{2\cdot (5!)^2}\epsilon^{a b_1 \ldots b_5 c_1 \ldots c_5}B_{b_1 \ldots b_5}\wedge B_{c_1 \ldots c_5}-   \frac{\rm i}{2}\bar{\psi} \wedge \Gamma^a \psi =0\, ,\\
& D B^{a_1 \ldots a_5} - 5 e B^{m [a_1}\wedge B^{a_2 \ldots a_5 ]}_{\; \; \; \; \; \; \; \;\;\;\; \; m}- \frac{e}{5!}\epsilon^{a_1 \ldots a_5 b_1 \ldots b_6}B_{b_1 \ldots b_5}\wedge V_{b_6}+ \\
&+\frac{5 e}{6!}\epsilon^{a_1 \ldots a_5 b_1 \ldots b_6}B^{c_1 c_2}_{\;\;\;\;\;\;\; b_1 b_2 b_3}\wedge B_{c_1 c_2 b_4 b_5 b_6}- \frac{\rm i}{2}\bar{\psi}\wedge \Gamma^{a_1 \ldots a_5}\psi =0\, ,\\
& D \psi - \frac{\rm i}{2} e \Gamma_a \psi \wedge V^a - \frac{1}{4}e \Gamma_{ab}\psi \wedge B^{ab}- \frac{\rm i}{2 \cdot 5!} e \Gamma_{a_1 \ldots a_5}\psi \wedge B^{a_1 \ldots a_5}=0\,.
\end{aligned}
\end{equation}

If one requires, as an extra condition, that the Lorentz $\so(1,10)$ spin connection $\omega^{ab}$  satisfies:
\begin{equation}
\mathcal{R}^{ab}=d\omega^{ab} -\omega^{ac}\wedge \omega_c^{\ b}=0 , \label{min}
\end{equation}
corresponding to a Minkowski background $D^2=0$,
then the first equation in (\ref{osp32'}), which corresponds to the Maurer-Cartan equation for the $\osp(1|32)$ connection, splits into two equations, namely equation (\ref{min}) plus the following condition
\begin{equation}\label{bab}
DB^{ab} + e B^{ac}\wedge B_{c}^{\ b}  = e V^a \wedge V^b+ \frac{e}{4 !}B^{a b_1 \ldots b_4}\wedge B^b_{\;\; b_1 \ldots b_4} +\frac{1}{2}\bar{\psi}\wedge \Gamma^{ab}\psi\,,
\end{equation}
defining the Maurer-Cartan equation for the new
 tensor field $B^{ab}$.

The algebra obtained from $\osp(1|32)$ through the above described procedure  is not isomorphic to $\osp(1|32)$ because of the extra constraint (\ref{min}), implying (\ref{bab}), which is imposed on (\ref{osp32'}). A slight generalization of it was introduced in the literature  in \cite{Castellani:1982kd} in 1982, soon after \cite{D'Auria:1982nx}, as a possible semisimple extension of the $DF$-algebra.
Actually, the algebra introduced in \cite{Castellani:1982kd} generalizes the algebra (\ref{osp32'}) with the constraint (\ref{min}), since it contains  an extra Maurer-Cartan equation for a spinor 1-form of length dimension 3/2. We will call it here $\eta_{SB}$ (to avoid confusion with the $\eta$ of the DF-algebra).
Its explicit form is:
\begin{equation}
\begin{aligned}
\mathcal{R}^{ab}&\equiv d\omega^{ab} -\omega^{ac}\wedge \omega_{c}^{\ b} = 0 , \\
DV^{a}&=- e B^{ab}\wedge V_b +  \frac{e}{2\cdot (5!)^2}\epsilon^{a b_1 \ldots b_5 c_1 \ldots c_5}B_{b_1 \ldots b_5}\wedge B_{c_1 \ldots c_5}+   \frac{\rm i}{2}\bar{\psi} \wedge \Gamma^a \psi , \\
D B^{ab} &=e V^a \wedge V^b- e B^{ac}\wedge B_{c}^{\; b}+ \frac{e}{24}B^{a b_1 \ldots b_4}\wedge B^b_{\;\; b_1 \ldots b_4}+ \frac{1}{2}\bar{\psi}\wedge \Gamma^{ab}\psi, \\
D B^{a_1 \ldots a_5} &=5 e B^{m [a_1}\wedge B^{a_2 \ldots a_5 ]}_{\; \; \; \; \; \; \; \;\;\;\; \; m} + \frac{e}{5!}\epsilon^{a_1 \ldots a_5 b_1 \ldots b_6}B_{b_1 \ldots b_5}\wedge V_{b_6}+ \\
\label{castella}& -\frac{5 e}{6!}\epsilon^{a_1 \ldots a_5 b_1 \ldots b_6}B^{c_1 c_2}_{\;\;\;\;\;\;\; b_1 b_2 b_3}\wedge B_{c_1 c_2 b_4 b_5 b_6}+ \frac{\rm i}{2}\bar{\psi}\wedge \Gamma^{a_1 \ldots a_5}\psi , \\
D \psi &= \frac{\rm i}{2} e \Gamma_a \psi \wedge V^a + \frac{1}{4}e \Gamma_{ab}\psi \wedge B^{ab}+ \frac{\rm i}{2 \cdot 5!} e \Gamma_{a_1 \ldots a_5}\psi \wedge B^{a_1 \ldots a_5}, \\
D \eta_{SB} &= \frac{\rm i}{2}   \Gamma_a \psi \wedge V^a + \frac{1}{4}  \Gamma_{ab}\psi \wedge B^{ab}+ \frac{\rm i}{2 \cdot 5!}  \Gamma_{a_1 \ldots a_5}\psi \wedge B^{a_1 \ldots a_5}= \frac 1 e D\psi\, ,
\end{aligned}
\end{equation}
where $D$, as before, denotes the Lorentz covariant derivative.
It is in fact a (Lorentz valued) central extension of (\ref{osp32'}) after imposing (\ref{min}), (\ref{bab}), since the dual of $\eta_{SB}$ is a nilpotent generator commuting with all the generators but the Lorentz ones, the rationale of its introduction being that of \emph{trying to reproduce} the DF algebra in the Inon\"{u}-Wigner contraction $e \rightarrow 0$. We shall refer to the algebra of \cite{Castellani:1982kd}, namely equations (\ref{castella}), as to the \emph{SB-algebra}, and to its semisimple subalgebra (\ref{osp32'}), (\ref{min}), (\ref{bab}) as to the \emph{restricted SB} algebra (RSB in the following). \footnote{The acronym SB(-algebra) stands for ``Stony Brook''(-algebra). The standard acronym referring to the names of the authors (CFGPV) would have been quite long. Having observed that the authors of Ref. \cite{Castellani:1982kd} were all affiliated to Stony Brook University, we found more convenient to adopt the shorter acronym SB.}

The algebra (\ref{castella}) is actually closed under differentiation even if one deletes the last equation containing the covariant differential $D\eta_{SB}$ (what corresponds to consider its subalgebra that we call here RSB algebra); This equation is in fact a double of the gravitino Maurer-Cartan equation, rescaled with the parameter $e$.
Furthermore, we see that the Maurer-Cartan equation for the 1-form $\eta_{SB}$ does not depend on any free parameter and, as such, in the limit $e\to 0$ \emph{it cannot be identified with the 1-form $\eta$} of the DF-algebra, see eq. (\ref{deta}).

We conclude
that at the price of introducing the (torsion) field $B^{ab}$ satisfying (\ref{bab}), the $\osp(1|32)$ algebra can be mapped into the RSB-algebra, whereby the spin connection $\omega^{ab}$ is identified with the Lorentz connection of a 11D Minkowski spacetime with vanishing Lorentz curvature (albeit with a modification of the (super)-torsion which is non-vanishing in both cases).
We shall refer to the RSB algebra also as ``torsion-deformed $\osp(1|32)$ algebra''.

The RSB-algebra  can be easily compared with the M-algebra,
 since its Maurer-Cartan equations have the virtue of reproducing exactly the $M$-algebra (but not the full \emph{DF-algebra}) by the Inon\"{u}-Wigner contraction $e \rightarrow 0$.

 For the RSB-algebra (\ref{osp32'}), (\ref{min}), (\ref{bab}), analogously to what happens   for the algebra $\osp(1|32)$ in the standard formulation (\ref{osp32}),  an interpretation in terms of ordinary superspace spanned by the supervielbein $(V^a,\psi)$ is not possible, because of the presence of two kinds of extra ``vielbeins" $B^{ab}$ and $B^{a_1\cdots a_5}$, whose dual generators are not (Lorentz-valued) central charges in this case.
 Indeed the bosonic  1-forms  $B^{ab}$ and $B^{a_1\cdots a_5}$ are elements of a \emph{semisimple} bosonic subalgebra  and as such, independently of their super-extension, they cannot be related to central charges.
  The same observation   also holds for the SB-algebra since it shares the same bosonic subalgebra with the RSB algebra.

 On the other hand, the DF Lie superalgebra, together with its bosonic subalgebra, is \emph{non-semisimple} and it enjoys a fiber bundle structure over ordinary superspace, where the fiber includes, besides the Lorentz connection, also the 1-forms $B^{ab}$ and $B^{a_1\cdots a_5}$, which in this theory are dual to Lorentz-valued central charges and can therefore be interpreted as abelian gauge fields on superspace \cite{Andrianopoli:2016osu}.
  
  At the dynamical level,  the space-time components $B^{ab}_{\;\;\;|c}$, $B^{a_1\cdots a_5}_{\;\;\;\;\;\;\;\;\;\;|c}$  of the 1-form gauge fields $B^{ab}$, $B^{a_1\cdots a_5}$ (we are using rigid Lorentz indices)  have extra degrees of freedom with respect to the component fields $A_{[abc]}$ \footnote{The possible interpretation of the field $A_{\mu\nu\rho}$ of 11D supergravity in terms of the totally antisymmetric part of the contorsion tensor in $\osp(1|32)$ was already considered in \cite{Troncoso:1997va}.
} and  $B_{[a_1\cdots a_6]}$, appearing in the FDA
on which 11D supergravity is based \footnote{$B_{[a_1\cdots a_6]}$ are the components of the 6-form $B^{(6)}$, related to $A^{(3)}$ by Hodge-duality of their field-strengths.}. As we are going to clarify in Section \ref{relating}, the extra degrees of freedom are dynamically decoupled from the physical spectrum in the DF algebra (contrary to what happens for the M-algebra) because of the presence of the nilpotent spinor generator $\eta$, which thus behaves as a BRST-ghost guaranteeing the equivalence of the hidden algebra with the super FDA. \footnote{This mechanism does not work for the semisimple RSB-algebra, since in that case the extra components in $B^{ab}_{\;\;\;|c}$, $B^{a_1\cdots a_5}_{\;\;\;\;\;\;\;\;\;\;|c}$ besides the fully antisymmetrized ones  are not decoupled from the physical spectrum.}

The detailed relation of the full SB-algebra with the DF-one (including the relation and differences between the nilpotent spinors $\eta_{SB}$ and $\eta$ of the two algebras) is more subtle and will be analyzed in the following section.

We conclude this section by analyzing some properties of the RSB-superalgebra related to its feature of being a semisimple superalgebra.

For semisimple Lie algebras, as it is well known from the Chevalley-Eilenberg cohomology of Lie algebras, and as already pointed out in \cite{Castellani:1982kd}, it is always possible to define a non-trivial 3-cocycle $H^{(3)}$ ($dH^{(3)}=0$) given by:
\begin{equation}\label{3cocycle}
 H^{(3)}= C_{ABC}\sigma^A\wedge\sigma^B\wedge\sigma^C = -2 h_{AB}\sigma^A \wedge d \sigma^B\, ,
\end{equation}
where $C_{ABC}=h_{AL}C^L_{\ BC}$ are the structure constants of the algebra, with an index lowered with the Killing metric $h_{AB}$.
The closure of $H^{(3)}$ is easily proven by using the Maurer-Cartan equations:
\begin{equation}\label{MC}
  d\sigma^A+\frac12 C^A_{BC}\sigma^B\wedge\sigma^C=0 ,
\end{equation}
where the $\sigma^A$ 1-forms are in the coadjoint representation of the (super)-Lie algebra. Indeed:
\begin{equation}\label{ji}
  dH^{(3)}=- \frac 32 C_{ABC}C^C_{\ LM}\sigma^A\wedge\sigma^B\wedge\sigma^L\wedge \sigma^M=0 \,,
\end{equation}
its vanishing being due to Jacobi identities.

For the case of the semisimple RSB-algebra, the set of 1-forms correspond to $\sigma^A=\{\omega^{ab},V^a,\psi^\alpha,\- B^{ab},B^{a_1\cdots a_5}\}$ . However, the Lorentz quotient of the RSB group admits the Lorentz-covariant Maurer-Cartan equations:
\begin{equation}\label{MC2}
  D\sigma^\Lambda+\frac12 C^\Lambda_{\ \Sigma\Gamma}\sigma^\Sigma\wedge\sigma^\Gamma=0
\end{equation} for the restricted set of 1-forms $\sigma^\Lambda=\{V^a,\psi^\alpha,B^{ab},B^{a_1\cdots a_5}\}$, allowing to rewrite, in this case:
\begin{equation}\label{h3h}
  H^{(3)}=-2\sigma^\Lambda\wedge  D\sigma^\Sigma h_{\Lambda\Sigma}\,,
\end{equation}
and satisfying $dH^{(3)}=0$.
From direct calculation we find, up to overall normalization, that the cocycle $H^{(3)}$ can be written as:
\begin{eqnarray}\label{3cocycle2}
 H^{(3)}&=& V^a\wedge DV_a +\frac 12 B^{ab}\wedge DB_{ab}+\frac 1{5!} B^{a_1 \cdots a_5}\wedge DB_{a_1 \cdots a_5}-  \frac 1 e \bar\psi \wedge D\psi\label{h1}\\
 &=&
 e\Bigl( B_{ab} \wedge V^a \wedge V^b + \frac 13 B_{a b}\wedge B^{b} _{\;c}\wedge B^{c a}+\frac 1{4!}B^{b_1 b_2} \wedge  B_{b_1 a_1...a_4}\wedge B_{b_2}^{\; \ a_1...a_4}+ \nonumber\\
& & + \frac 1{(5!)^2}  \epsilon_{a_1...a_5 b_1...b_5 m}B^{a_1...a_5}\wedge B^{b_1...b_5}\wedge V^m + \nonumber\\
& &- \frac 13 \frac 1{[2!\cdot (3!)^2 \cdot 5!]}  \epsilon_{m_1...m_6 n_1...n_5}B^{m_1m_2m_3p_1p_2}\wedge B^{m_4m_5m_6p_1p_2}\wedge B^{n_1...n_5}\Bigr)\,.\label{h2}
 \end{eqnarray}
 We observe that it is actually a bosonic 3-form, see eq. (\ref{h2}),  the same expression  holding for the 3-cocycle of its bosonic subalgebra.

An analogous result  for $\osp(1|32)$ can be obtained by setting $B^{ab}= 0$ in (\ref{h1}), (\ref{h2}).

Let us remark that the $e\to 0$ limit of $H^{(3)}$ is a singular limit: $H^{(3)}\to 0$, but $\frac 1e H^{(3)}$ is finite if one considers the second expression (\ref{h2}), while $\frac 1e d H^{(3)}\neq 0$ in the limit, corresponding to the fact that the Killing metric of the contracted superalgebra at $e\to 0$ is degenerate.

For $e\neq 0$, instead, $ H^{(3)}$ is a 3-cocycle of the superalgebra and, following the general Sullivan construction of FDAs \cite{Sullivan} (for a review, see for example \cite{Castellani:1991et}), it could be trivialized in terms of a 2-form $Q^{(2)}$ writing:
\begin{equation}\label{dq}
  dQ^{(2)}+H^{(3)}=0,
\end{equation}
thus realizing a new FDA in the semisimple case.

It could be interesting to investigate about a hidden superalgebra of (\ref{dq}), which would allow to parametrize $Q^{(2)}$ in terms of an appropriate set of 1-form generators.
However, to ascertain whether one can associate a hidden super-Lie algebra to the FDA (\ref{dq}) one has to introduce extra fields besides the set of generators $\{\sigma^\Lambda\}$ of the SB-algebra.
This is left to a future investigation.

\section{Relating $\osp(1|32)$ to DF-algebra}\label{relating}

We would like to clarify here the relation between the DF-algebra of \cite{D'Auria:1982nx} and the SB-algebra that, as described in Section \ref{unusual}, is a Lorentz valued central extension of the $\osp(1|32)$ algebra in a  torsion deformed form in which the Lorentz spin connection is decomposed into a flat spin connection and an appropriate tensor 1-form $B^{ab}$.

To this aim, let us first write, from \cite{D'Auria:1982nx,Andrianopoli:2016osu}, the Maurer-Cartan equations for the 1-forms $\sigma^\Lambda=\{V^a,\psi_\alpha,B^{ab},B^{a_1\cdots a_5},\eta_\alpha\}$ of the DF-algebra (in their Lorentz-covariant formulation):
\begin{equation}
\begin{aligned}
R^{ab}&\equiv d\omega^{ab} - \frac 12 \omega^{ac}\wedge \omega_{c}^{\ b}=0\,,\\
 D V^a &= \frac{\ii}{2}\overline{\Psi}\wedge \Gamma^a \Psi  \,,\label{DFalg} \\
D \Psi&=0\,,\\
D B^{a_1a_2} & =  \frac{1}{2}\overline{\Psi}\wedge \Gamma^{a_1a_2}\Psi , \\
D B^{a_1...a_5}& =  \frac{\ii}{2} \overline{\Psi}\wedge \Gamma^{a_1...a5}\Psi\,,\\
D \eta & =  \ii E_1 \Gamma_a \Psi \wedge V^a + E_2 \Gamma_{ab}\Psi \wedge B^{ab}+ \ii E_3 \Gamma_{a_1...a_5}\Psi \wedge B^{a_1...a_5}\,.
\end{aligned}
\end{equation}
Integrability of $D \eta $-equation implies, using Fierz identities of the 1-forms $\psi$ in superspace, the following relation among the parameters $E_i$:
\begin{equation}\label{relat}
  E_1+10 E_2-6!E_3=0.
\end{equation}
Since one of the $E_i$ 's can be reabsorbed in the normalization of $\eta$, the DF-algebra depends on one free parameter, as it was pointed out in \cite{Bandos:2004xw}.

As shown in \cite{D'Auria:1982nx}, the DF-algebra 1-forms spanning the Maurer-Cartan equations allow to express the 3-form $A^{(3)}$ of the FDA of 11D supergravity in terms of 1-forms $\sigma^\Lambda$, which, in the notations of \cite{Andrianopoli:2016osu} is:
\begin{eqnarray}\label{29}
A^{(3)}(\sigma) & = & T_0 B_{ab} \wedge V^a \wedge V^b + T_1 B_{a b}\wedge B^{b} _{\;c}\wedge B^{c a}+ \nonumber\\
& +& T_2 B_{b_1 a_1...a_4}\wedge B^{b_1}_{\; b_2}\wedge B^{b_2 a_1...a_4}+ T_3 \epsilon_{a_1...a_5 b_1...b_5 m}B^{a_1...a_5}\wedge B^{b_1...b_5}\wedge V^m + \nonumber\\
& +& T_4 \epsilon_{m_1...m_6 n_1...n_5}B^{m_1m_2m_3p_1p_2}\wedge B^{m_4m_5m_6p_1p_2}\wedge B^{n_1...n_5} + \nonumber\\
& + & \ii S_1 \overline{\Psi}\wedge \Gamma_a \eta \wedge V^a + S_2 \overline{\Psi}\wedge \Gamma^{ab}\eta \wedge B_{ab}+ \ii S_3 \overline{\Psi}\wedge \Gamma^{a_1...a_5}\eta \wedge B_{a_1...a_5}\,,\label{a3par}
\end{eqnarray}
where the real parameters $\{E_i, T_i,S_i\}$
can be written in terms of a single parameter, as clarified in \cite{Bandos:2004xw}, and
are given in \cite{Andrianopoli:2016osu}. For our purpose, it is convenient to observe that they can be rewritten as follows:
\begin{equation}\label{deco}
\left\{\begin{array}{ccl}
T_0 &=& \frac{1}{6}+ \alpha, \\
T_1 &=&-\frac{1}{90} + \frac{1}{3} \alpha, \\
T_2 &=&-\frac 1{4!} \alpha, \\
T_3 &=& \frac{1}{(5!)^2}\alpha , \\
T_4 &=& - \frac{1}{3[2!\cdot (3!)^2 \cdot 5!]} \alpha ,
\end{array}\right.
\quad
\left\{\begin{array}{ccl}
S_1 &=& \frac{1}{4!C} + \frac{1}{2\cdot (5!) E_3} \alpha, \\
S_2 &=& -\frac{1}{10\cdot (4!)C} + \frac{1}{4 \cdot(5!)E_3} \alpha, \\
S_3 &=& \frac{1}{2 \cdot (5!)^2 E_3} \alpha ,
\end{array}\right.
\quad
\left\{\begin{array}{ccl}
E_1 &=& -10 C + \frac{C^2}{E_3} \alpha, \\
E_2 &=& C + \frac{C^2}{2E_3} \alpha, \\
E_3 &=& \frac{C^2}{ 5! E_3} \alpha ,
\end{array}\right.
\end{equation}
where we have defined, using the notations of \cite{Andrianopoli:2016osu}, $C\equiv E_2-60 E_3$, $\alpha \equiv 5!\frac{E_3^2}{C^2}$.

Given the above expression, it is useful to decompose the 1-form spinor $\eta$ as follows:
\begin{eqnarray}
\eta= -10 C(\xi + \alpha \lambda), \label{etadec}
\end{eqnarray}
where we introduced the spinor 1-forms $\xi$ and $\lambda$ satisfying:
\begin{eqnarray}
D\xi &=&  \ii \Gamma_a \psi \wedge V^a -\frac 1{10} \Gamma_{ab}\psi \wedge B^{ab}\,,\label{dxi}\\
D\lambda &=&-\frac{C}{5 E_3}\left( \frac {\ii}2 \Gamma_a\psi \wedge V^a + \frac 14 \Gamma_{ab}\psi \wedge B^{ab} + \frac{\ii}{2(5!)}\Gamma_{a_1\cdots a_5}\psi \wedge B^{a_1\cdots a_5}\right)\nonumber\\
&=&-\frac{C}{5 E_3}D\eta_{SB}\,. \label{dlambda}
\end{eqnarray}
From eq. (\ref{dlambda}) we see that $\lambda$ can be chosen as proportional to the spinor 1-form $\eta_{SB}$ introduced in (\ref{castella}) as a Lorentz-valued central extension of the RSB-superalgebra:\\
 $\lambda=- \frac{C}{5 E_3}\eta_{SB}$.

eqs. (\ref{deco}) and (\ref{etadec}) allow to decompose also $A^{(3)}(\sigma)$ into two pieces, namely
\begin{equation}\label{a3d}
A^{(3)} = A^{(3)}_{(0)} + \alpha A^{(3)}_{(e)},
\end{equation}
where
\begin{equation}\label{a30}
A^{(3)}_{(0)}= \frac{1}{6}\left(B_{ab} \wedge V^a \wedge V^b - \frac{1}{15} B^{ab}\wedge B_{bc} \wedge B^c_{\;\; a} - \frac{5\ii}2\bar \psi \wedge \Gamma_a \xi \wedge V^a + \frac{1}{4} \bar \psi \wedge \Gamma_{ab} \xi \wedge B^{ab}\right)\,,
\end{equation}
while
\begin{equation}\label{a3e}
A^{(3)}_{(e)}= \frac 1 e H^{(3)} +2 \bar \eta_{SB}D \eta_{SB}\,,
\end{equation}
 where we recognize, in the first term in (\ref{a3e}), the  RSB-invariant 3-form $\frac 1e H^{(3)}$ introduced in (\ref{3cocycle2}), which in the $e\to 0$ limit is finite but looses its character of being a 3-cocycle (a closed form), becoming just a 3-cochain of the $M$-algebra. Explicitly we have:
\begin{eqnarray}\label{3cocyclelim}
\frac 1e H^{(3)}&=&
 \Bigl( B_{ab} \wedge V^a \wedge V^b + \frac 13 B_{a b}\wedge B^{b} _{\;c}\wedge B^{c a}+\frac 1{4!}B^{b_1 b_2} \wedge  B_{b_1 a_1...a_4}\wedge B_{b_2}^{\; \ a_1...a_4}+ \nonumber\\
& & + \frac 1{(5!)^2}  \epsilon_{a_1...a_5 b_1...b_5 m}B^{a_1...a_5}\wedge B^{b_1...b_5}\wedge V^m + \nonumber\\
& &- \frac 13 \frac 1{[2!\cdot (3!)^2 \cdot 5!]}  \epsilon_{m_1...m_6 n_1...n_5}B^{m_1m_2m_3p_1p_2}\wedge B^{m_4m_5m_6p_1p_2}\wedge B^{n_1...n_5}\Bigr)\,,
 \end{eqnarray}
 and, by straightforward differentiation using the Maurer-Cartan equations of DF-algebra (\ref{DFalg}), we easily verify that $d\left(\frac 1e H^{(3)}\right)_{e=0}\neq 0$, while:
\begin{eqnarray}
dA^{(3)}_{(0)}&=&\frac 12 \bar \psi \wedge \Gamma_{ab} \psi \wedge V^a \wedge V^b\label{da30},\\
dA^{(3)}_{(e)}&=&0\label{da3e}\,.
\end{eqnarray}
Some remarks are in order. First of all, let us observe that $A^{(3)}_{(0)}$ only depends on the restricted set of 1-forms $\{V^a,\psi,B^{ab},\xi\}$, not including $B^{a_1\cdots a_5}$,
 through an expression, (\ref{a30}), which \emph{does not contain any free parameters}. The term $A^{(3)}_{(0)}$ is however the only one contributing to the (vacuum) 4-form cohomology in superspace, eq. (\ref{da30}), $A^{(3)}_{(e)}$ being instead a closed 3-form in the vacuum. \footnote{Surprisingly, it corresponds to one of the solutions found in the original paper of D'Auria-Fr\'{e} \cite{D'Auria:1982nx}.}

On the other hand, we see that the 1-parameter family of solutions to the DF-algebra, whose presence was clarified in \cite{Bandos:2004xw}, actually only depends on the contribution $A^{(3)}_{(e)}$, which appears as a trivial deformation of $A^{(3)}_{(0)}$ in $A^{(3)}$, since it does not contribute to the vacuum 4-form cohomology (\ref{solito}).
$A^{(3)}_{(e)}$ is however invariant not
only under the DF-algebra (\ref{DFalg}), but also under the SB-algebra, \emph{even at finite $e$}. The other term $A^{(3)}_{(0)}$ does instead explicitly breaks the invariance under the SB-algebra.

Let us spend some words to discuss the role of the spinor 1-forms $\xi$ and $\lambda$ introduced in the decomposition (\ref{etadec}) of $\eta$, and appearing in $A^{(3)}$ (\ref{a3d}), following the lines of the discussion given in \cite{Andrianopoli:2016osu}: The spinor $\xi$ appears in $A^{(3)}_{(0)}$, and its role is to allow for $dA^{(3)}_{(0)}$ to be a closed 4-form \emph{on ordinary superspace}; It behaves as a cohomological ghost, since its supersymmetry and gauge transformations exactly cancel the  non-physical contributions from $B^{ab}$. The group manifold generated by the set of $\{\sigma^\Lambda\}$ including
 $\xi$ has a fiber-bundle structure with ordinary superspace as base space \cite{Andrianopoli:2016osu}.

As for the second spinor, $\lambda \propto \eta_{SB}$, appearing instead in the $\osp(1|32)$ invariant term $A^{(3)}_{(e)}$, at first sight its role could appear less clear, since $dA^{(3)}_{(e)}=0$ in the  FDA where the vacuum relation  (\ref{solito}) holds. It plays however a role analogous to the one of $\xi$: Indeed, in the absence of its contribution, $A^{(3)}_{(e)}$ would reduce to the bosonic 3-form $\frac 1 e H^{(3)}$ that, as we already observed at the end of Section \ref{unusual}, is a closed 3-form for $e\neq 0$, this property being lost in the limit $e\to 0$. In the same limit, $ \frac 1 e d H^{(3)}$ is instead a 4-form polynomial of all the $\sigma^\Lambda$, that is a cochain of the enlarged superspace including $B^{ab}$ and $B^{a_1\cdots a_5}$. The role of $\eta_{SB}$ is then crucial to restore, also for $\alpha \neq 0$,
the correct 4-form cohomology (\ref{solito}) on the vacuum superspace for $dA^{(3)}$, by allowing $dA^{(3)}_{(e)}=0$. \footnote{On the other hand, in the semisimple case $e\neq 0$, $H^{(3)}$ is a closed 3-form and $\eta_{SB}$ looses its cohomological role.}

We remark that considering the interacting theory out of the vacuum, one should introduce a 4-form super fieldstrength $G^{(4)}$ in superspace:
\begin{equation}\label{g4}
 G^{(4)}\equiv dA^{(3)}-\frac12 \bar\psi\wedge\Gamma_{ab}\psi\wedge V^a\wedge V^b\,.
\end{equation}
In this case one would expect that the superspace 4-form cohomology could also receive non-trivial contributions from $dA^{(3)}_{(e)}$.

\section{Conclusions}

We have found that, despite of the fact that the $M$-algebra is a In\"on\"u-Wigner contraction of the $\osp(1|32)$ algebra \footnote{More precisely, of its torsion deformation described in Section \ref{unusual} (what we called RSB-algebra).},
 still the DF-algebra cannot be obtained as In\"on\"u-Wigner contraction from the SB-algebra that, as we discussed, is a (Lorentz-valued) central extension of the RSB-algebra.
  Correspondingly, 11D supergravity is  not left invariant by the $\osp(1|32)$ algebra (not even in its torsion deformed formulation RSB), while being invariant under the DF-superalgebra.
  This is due to the fact that the spinor 1-form $\eta$ of the DF-algebra (that, as we have discussed, is a  spinor ``central'' extension of the $M$-algebra) contributes to the DF-algebra with structure constants different from  the ones of the SB-algebra  (which is   related to the $\osp(1|32)$ algebra, as discussed in Section \ref{unusual}). In particular, referring to eq. (\ref{etadec}), we see that $\eta$ differs from ${\eta_{SB}}\propto \lambda$ by the extra 1-form generator $\xi$.

 This has a counterpart in the expression of $A^{(3)}= A^{(3)}(\sigma^\Lambda)$, which trivializes the vacuum 4-form cohomology in superspace in terms of DF-algebra 1-form generators $\sigma^\Lambda$.
 As the decomposition (\ref{a3d}) shows, $A^{(3)}(\sigma^\Lambda)$ is not invariant under the $\osp(1|32)$ algebra (neither under its torsion deformation RSB) because of the contribution $A^{(3)}_{(0)}$, explicitly breaking this symmetry.
 Such term is however the only one contributing to the vacuum 4-form cohomology in superspace, due to the presence in the DF-algebra of the two spinors $\xi$ and $\eta_{SB}$ into which the cohomological spinor $\eta$ can be decomposed.

 A still open problem is to perform a similar analysis for the 6-form $B^{(6)}$ of 11D supergravity. As we discussed in \cite{Andrianopoli:2016osu}, we expect in this case that a cohomological 1-form spinor different from $\eta$ should play a crucial role. The decomposition (\ref{etadec}) of $\eta$  into a linear combination of 1-form spinors, $\xi$ and $\eta_{SB}$, suggests that possibly the relevant spinor in the case of $B^{(6)}$ could correspond to a linear combination of $\xi$ and $\eta_{SB}$ different from (\ref{etadec}). Such analysis should preliminarly require the knowledge of the parametrization of $B^{(6)}$ in terms of 1-forms, which is not available yet.

 The above decomposition of $A^{(3)}(\sigma^\Lambda)= A^{(3)}_{(0)}+ \alpha  A^{(3)}_{(e)}$ in superspace, where we disclosed   different contributions to the 4-form cohomology on superspace from the two terms $dA^{(3)}_{(0)}(\sigma^\Lambda)$ and  $dA^{(3)}_{(e)}(\sigma^\Lambda)$, suggests that the above contributions could be possibly related to the general analysis of \cite{Witten:1996md, Diaconescu:2000wy, Diaconescu:2003bm}, where the 4-form cohomology of $M$-theory on a spin manifold $Y$ is shown to be shifted, with respect to the integral cohomology class, by the canonical integral class of the spin bundle of $Y$. Referring to (\ref{g4}), it appears reasonable to conjecture that one could rephrase the above statement into the following one, in terms of the super field-strength  $G^{(4)}$ in superspace:  $G^{(4)}$ has integral periods in superspace, while  the periods of $dA^{(3)}$ are shifted by the contribution (possibly fractional) of the spin bundle. Since our analysis refers to the FDA describing the vacuum
 in  \emph{superspace} we should consider,  as spin manifold  $Y$, flat superspace, where the \emph{integral} cohomology class is trivial.    This corresponds, in our formulation, to the trivial contribution from the RSB-invariant term $A^{(3)}_{(e)}(\sigma^\Lambda)$, the only non-trivial contribution to the 4-form cohomology on flat superspace coming from   $dA^{(3)}_{(0)}(\sigma^\Lambda)$, which accounts for  the contribution from the spin bundle.  A deeper analysis of the correspondence between the two approaches, for the vacuum theory and for the dynamical theory out of the vacuum, is currently under investigation and left to future work. In particular, it is still to be explicitly shown that the contribution to the 4-form cohomology in superspace from  $dA^{(3)}_{(0)}(\sigma^\Lambda)$ could assume both integer and half-integer values. In this direction, the techniques developed in \cite{Castellani:2016ibp}, where a formulation of supergravity in superspace with integral forms was introduced, could be particular useful.

It appears that the nilpotent spinor 1-form $\eta$ could be an important addition towards the construction of a possible off-shell theory underlying 11D supergravity.
In \cite{Hassaine:2003vq}, a supersymmetric 11D lagrangian invariant under the $M$-algebra and closing off-shell without requiring auxiliary fields was constructed, as a Chern-Simons form. It would be very interesting to investigate the possible connections between the two approaches.

Finally, let us stress that the description of 11D supergravity in terms of its hidden DF-algebra could be particularly useful in the analysis of its compactification to lower dimensions: The 1-form fields $\sigma^\Lambda$ of the DF-algebra should give an alternative description of exceptional field theory (see, for example \cite{Hohm:2013pua, Hohm:2013uia, Hohm:2014qga} and references therein) where the section constraints, required in that theory to project the field equations on ordinary superspace, should be dynamically implemented through the presence of the cohomological spinor $\eta$. Some work is in progress on this topic \cite{wip}.

\section*{Acknowledgements}
We are grateful to Bianca Letizia Cerchiai, Mario Trigiante and Jorge Zanelli for stimulating discussions and comments. L.A. and R.D'A. also acknowledge an interesting discussion with Paolo Aschieri, Andreas Deser, Leonardo Castellani and Christian Saemann.
L.A. would like to thank the Galileo Galilei Institute for Theoretical Physics for
the kind hospitality during the early stages of the preparation of this work.

\end{document}